# MLatom 3: Platform for machine learning-enhanced computational chemistry simulations and workflows

Pavlo O. Dral,[1,2*] Fuchun Ge,[1,2] Yi-Fan Hou,[1,2] Peikun Zheng,[1,2] Yuxinxin Chen,[1,2] Mario Barbatti,[3,4] Olexandr Isayev,[5] Cheng Wang,[1,6] Bao-Xin Xue,[1,2,7] Max Pinheiro Jr,[3,8] Yuming Su,[1,6] Yiheng Dai,[1,6,9] Yangtao Chen,[1,6] Lina Zhang,[1,2] Shuang Zhang,[1,2,10] Arif Ullah,[11] Quanhao Zhang,[1,2] Yanchi Ou[1,2,12]

[1]*State Key Laboratory of Physical Chemistry of Solid Surfaces, College of Chemistry and Chemical Engineering, and Innovation Laboratory for Sciences and Technologies of Energy Materials of Fujian Province (IKKEM), Xiamen University, Xiamen, Fujian 361005, China*

[2]*Fujian Provincial Key Laboratory of Theoretical and Computational Chemistry, Xiamen, Fujian 361005, China*

[3]*Aix Marseille University, CNRS, ICR, 13013 Marseille, France*

[4]*Institut Universitaire de France, 75231, Paris, France*

[5]*Department of Chemistry, Carnegie Mellon University, Pittsburgh, Pennsylvania 15213, USA*

[6]*iChem, Xiamen University, Xiamen, Fujian 361005, China*

[7]*Present address: Xiamen Double Ten Middle School, Xiamen, Fujian 361009, China*

[8]*Present address: Alstom Transport S.A., Saint-ouen-sur-seine, France*

[9]*Present address: Beijing National Laboratory for Molecular Sciences, College of Chemistry and Molecular Engineering, Peking University, Beijing 100190, China*

[10]*Present address: Neotrident (Suzhou) Co., Ltd., Suzhou, Jiangsu 215028, China*

[11]*School of Physics and Optoelectronic Engineering, Anhui University, Hefei 230601, China*

[12]*Present address: Shanghai Mayoo Technology Inc, Shanghai 201318, China*

*Email: dral@xmu.edu.cn. Web: http://dr-dral.com*

**Abstract**

Machine learning (ML) is increasingly becoming a common tool in computational chemistry. At the same time, the rapid development of ML methods requires a flexible software framework for designing custom workflows. MLatom 3 is a program package designed to leverage the power of ML to enhance typical computational chemistry simulations and to create complex workflows. This open-source package provides plenty of choice to the users who can run simulations with the command line options, input files, or with scripts using MLatom as a Python package, both on their computers and on the online XACS cloud computing at XACScloud.com. Computational chemists can calculate energies and thermochemical properties, optimize geometries, run molecular and quantum dynamics, and simulate (ro)vibrational, one-photon UV/vis absorption, and two-photon absorption spectra with ML, quantum mechanical, and combined models. The users can choose from an extensive library of methods containing pre-trained ML models and quantum mechanical approximations such as AIQM1 approaching coupled-cluster accuracy. The developers can build their own models using various ML algorithms. The great flexibility of MLatom is largely due to the extensive use of the interfaces to many state-of-the-art software packages and libraries.

# 1 Introduction

Computational chemistry simulations are common in chemistry research thanks to abundant general-purpose software, most of which has started as purely quantum mechanical (QM) and molecular mechanical (MM) packages. More recently, the rise of artificial intelligence (AI)/machine learning (ML) applications for chemical simulations has caused the proliferation of programs mostly focusing on specific ML tasks such as learning potential energy surfaces (PESs).[1-17] The rift between the development of the traditional QM and MM packages on the one hand and ML programs on the other hand, is bridged to some extent by the higher-level library ASE,[18] which enables usual computational tasks via interfacing heterogeneous software. The further integration of QM, MM, and ML has been prompted by the maturing of ML techniques and is evidenced by the growing trend of incorporating ML methods in the QM and MM computational chemistry software.[13, 19-21]

Against this backdrop, the MLatom package started in 2013 as a pure stand-alone ML package to provide a general-purpose experience for computational chemists akin to the black-box QM packages.[22] The early MLatom could be used for training, testing, and using ML models and their combinations with QM methods (e.g., Δ-learning[23] and learning of Hamiltonian parameters[24]), accurate representation of PES,[25, 26] sampling of points from data sets,[26] ML-accelerated nonadiabatic dynamics,[27] and materials design[28]. The fast pace of method and software development in QM, MM, ML, and other computational science domains led to MLatom 2 which started to include interfaces to third-party packages.[29] Such an approach provided a unique opportunity for the package users to choose one of the many established ML models – similar to the users of the traditional QM software who can choose one of the many QM methods. MLatom 2 could perform training of the ML models, evaluate their accuracy, and then use the models for geometry optimizations and frequency calculations. Special workflows were also implemented, such as acceleration of the absorption UV/vis spectra calculations with ML[30] and prediction of two-photon absorption spectra[31]. In addition, MLatom 2 could be used to perform simulations with general-purpose AI-enhanced QM method[32] AIQM1 and universal machine learning potentials of the ANI family[2, 33-35] with the accurate scheme developed for calculating heats of formation[36] with uncertainty quantification with these methods.

With time, the need to develop increasingly complex workflows that incorporate ML and QM for a broad range of applications has necessitated the rethink and redesign of MLatom to enable the rapid development of highly customized routines. These additional design requirements for MLatom to serve not just as a black-box general-purpose package but also as a flexible platform for developers resulted in a significant extension, redesign, and rewrite of the program. The subsequent upgrade has allowed the use of MLatom through the versatile Python API (MLatom PyAPI) and also included the implementation of more simulation tasks such as molecular and quantum dynamics and the support of QM methods and composite schemes based on the combinations of QM and ML models. This upgrade was released[37] as MLatom 3 in 2023 – ten years after the start of the project. During this decade, MLatom went through a drastic transformation from a pure Fortran package to a predominantly Python package with one-third of the code written in Fortran for efficient implementations of critical parts. MLatom 3 comes under the open-source permissive MIT license (modified to request proper citations). Here we give an overview of the capabilities of MLatom 3 and provide examples of its applications.

## 2 Overview

**Features**

**Models**

- AI-enhanced QM methods and pre-trained ML models (AIQM1, ANI-1ccx, etc.)
- **V3** simulations with QM methods
- simulations with user-defined models

**Simulations**

- single-point calculations
- geometry optimizations (minima, TS, IRC)
- frequencies & thermochemistry
- **V3** molecular dynamics
- **V3** quantum dynamics with machine learning
- **V3** IR and power spectra from MD
- UV/vis spectra (ML-NEA)
- two-photon absorption cross sections (ML-TPA)

**Learning**

- training popular ML models (KREG, ANI, sGDML, PhysNet, DPMD, GAP-SOAP, KRR-CM)
- training generic ML models (kernel ridge regression with many kernels)
- multi-component models (Δ-learning, self-correction)
- optimizing hyperparameters
- evaluating ML models (also with learning curves)

**Data**

- databases
- molecular descriptors (RE, Coulomb matrix, …)
- analyzing data sets
- splitting data sets (training, validation…)
- sampling (random, structure-based, farthest-point)

**V3 MLatom Python API**

- data in different formats (Python class objects, json, h5) and different types (molecules, trajectories)
- models
  - methods that do not need training (QM methods, pre-trained universal ML models and combined QM/ML models)
  - models designed by the user
    - one of the many standard architectures (ANI, PhysNet, DeepPot-SE, KREG, sGDML, GAP-SOAP)
    - other custom models including those based on combinations of methods and models
- access to simulation tasks based on models and their application to molecules
  - single-point calculations, geometry optimizations, frequency and thermochemical calculations, molecular dynamics
  - parallelization of tasks, finite-difference derivatives

* 'V3' marks implementations released in MLatom 3

**Figure 1.** Overview of the MLatom 3 capabilities.

MLatom merges the functionality from typical quantum chemical and other atomistic simulation packages and the capabilities of desperate ML packages with a strong focus on molecular systems. The user can choose from a selection of ready-to-use QM and ML models and design and train ML models to perform the required simulations. The bird's view of the MLatom capabilities is best given in Figure 1.

One of the current main goals of MLatom is to enable simulation tasks of interest for a computational chemist with generic types of models that can be based on ML, QM, and their combinations (see Section 4). These tasks include single-point calculations, optimization of geometries of minima and transition states (which can be followed by intrinsic reaction coordinate (IRC) analysis[38]), frequency and thermochemical property calculations, molecular and quantum dynamics, rovibrational (infrared (IR) and power) spectra, ML-accelerated UV/vis absorption and two-photon absorption spectra simulations. This part of MLatom is more similar to traditional QM and MM packages but with much more flexibility in model choice and unique tasks. A dedicated Section 5 will give a more detailed account of the simulations.

Enabling the users to create their own ML models was MLatom's original main focus and it continues to play a major role. The MLatom supports a range of carefully selected representative ML algorithms that can learn the desired properties as a function of a 3D atomistic structure. Typically, these algorithms are used, but not limited to, for learning PESs and hence often can be called, for simplicity, ML (interatomic) potentials (MLPs).[39-43] One particular specialization of MLatom is the original implementation of kernel ridge regression (KRR) algorithms for learning any property as a function of any user-provided input vectors or XYZ molecular coordinates.[22] In addition, the user can create custom multi-component models based on concepts of Δ-learning,[23] hierarchical ML,[25] and self-correction[26]. These models may consist of ML and QM methods. MLatom provides standardized means for training, hyperparameter optimization, and evaluation of the models so that switching from one model type to another may need just one keyword change.[29] This allows one to easily experiment with different models and choose the most appropriate for the task.

The data is as important as choosing and training the ML algorithms. MLatom 3 provides several data structures specialized for computational chemistry needs, mainly based on versatile Python classes for atoms, molecules, molecular databases, and dynamics trajectories. These classes allow not just storing the data in a clearly structured format, but also handling it by, e.g., converting to different molecular representations and data formats and splitting and sampling the data sets into the training, validation, and test subsets. Because data is a central concept in the age of data-driven models and MLatom as a package, we describe data structures in Section 3 before describing models, simulations, and machine learning.

How the user interacts with the program is also important and ideally the features should be easily accessible and their use intuitive. MLatom calculations can be requested by providing command-line options either directly or through the input file. Alternatively, MLatom can be used as a Python module which can be imported and used for creating calculation workflows of varying complexity. A side-by-side comparison of these two approaches is given in Figure 2. More examples highlighting different use cases of MLatom are interspersed throughout this article.

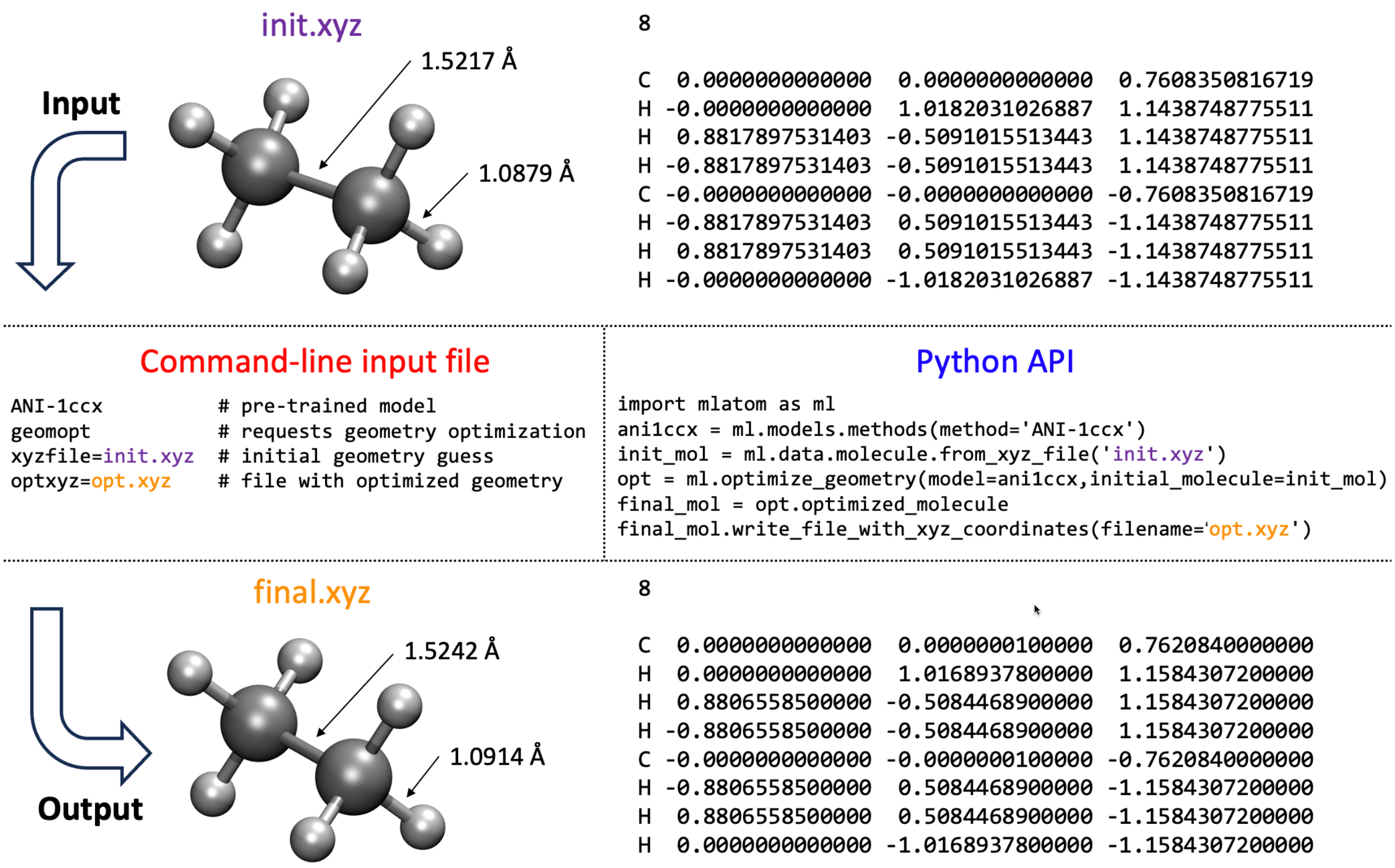

**Figure 2.** Side-by-side comparison of the usage of MLatom in both command-line mode and via Python API for a common task of geometry optimization with one of the pre-trained ML models ANI-1ccx.

MLatom as an open-source package can be conveniently installed via PyPI, i.e., simply using the command `pip install mlatom` or from the source code available on GitHub at https://github.com/dralgroup/mlatom. To additionally facilitate access to AI-enhanced computational chemistry, MLatom can be conveniently used in the XACS cloud computing service at https://XACScloud.com whose basic functionality is free for non-commercial uses such as education and research. Cloud computing eliminates the need for program installation and might be particularly useful for users with limited computational resources.

## 3 Data

In MLatom, everything revolves around operations on data: databases and data points of different types such as an atom, molecule, molecular database, and molecular trajectory (Figure 3). They are implemented as Python classes that contain many useful properties and provide different tools to load and dump these data-type objects using different formats. For example, the key type is a molecule that can be loaded from XYZ file or SMILES and then automatically parsed into the constituent atom objects. Atom objects contain information about the nuclear charge and mass as well as nuclear coordinates. A molecule object is assigned charge and multiplicity. The information about molecular and atomic properties can be passed to perform simulations, e.g., MD, with models that update and create new molecule objects with calculated quantum mechanical properties such as energies and energy gradients.

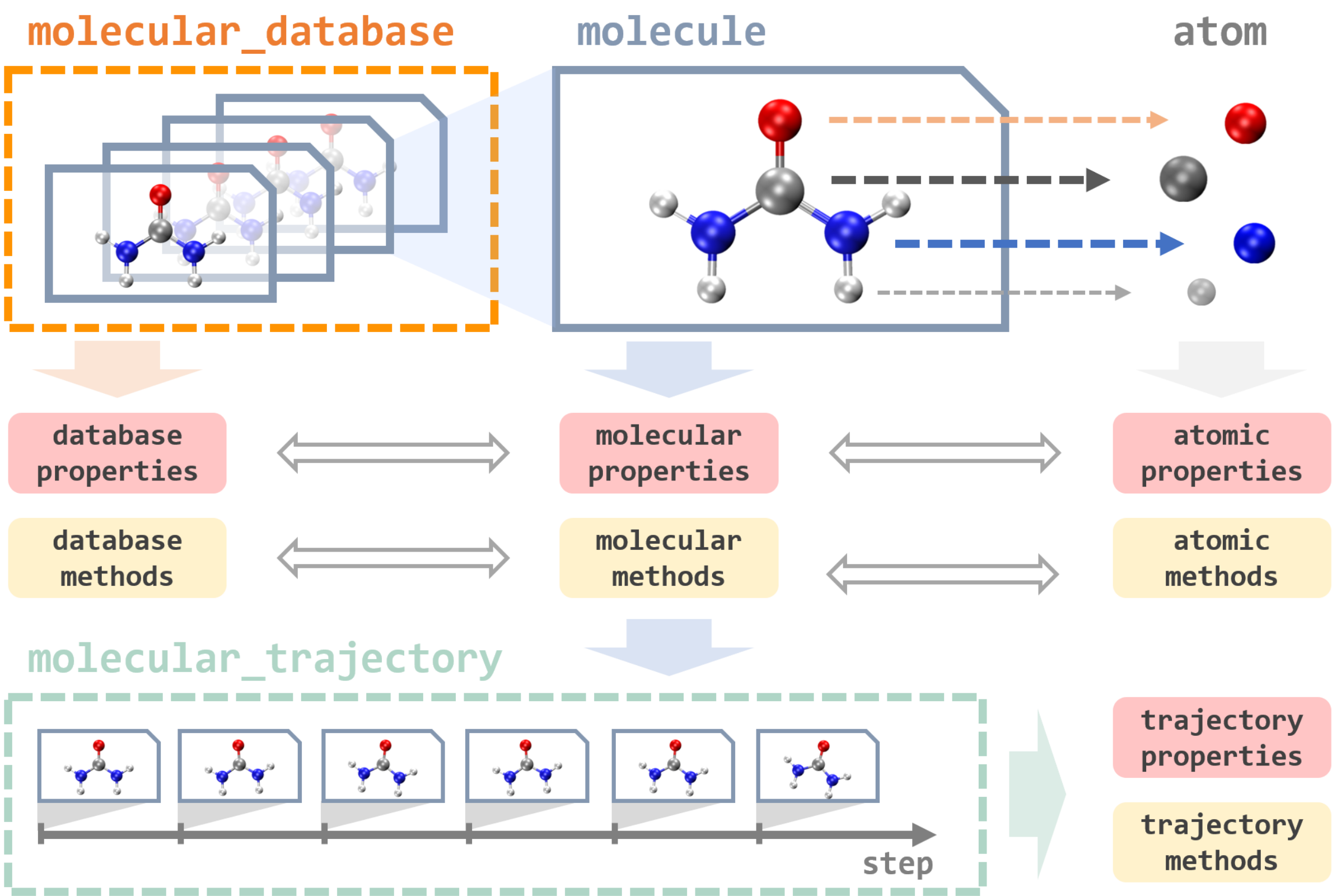

**Figure 3.** Overview of different data types in MLatom.

See Figure 2 for an example of loading a molecule object `init_mol` from the file `init.xyz`, used as the initial guess for the geometry optimization, returning an optimized geometry as a new molecule object `final_mol`, which is saved into the `final.xyz` file. Data objects can be directly accessed and manipulated via MLatom Python API. When using the MLatom in the command-line mode, many similar operations are done under the hood so that the user often just needs to prepare input files in standard formats such as files with XYZ coordinates.

Molecule objects can be combined into or created by parsing the molecular database that has, e.g., functions to split it into the different subsets needed for training and validation of ML models. The databases can be loaded and dumped in plain text (i.e., several files including XYZ coordinates, labels, XYZ derivatives), JSON, and npz formats. Another data type is molecular trajectory which consists of steps containing molecules and other information. Molecular trajectory objects are created during geometry optimization and MD simulations and in the latter case, the step is a snapshot of MD trajectory, containing information about the time, nuclear coordinates and velocities, atomic numbers and masses, energy gradients, kinetic, potential and total energies, and, if available, dipole moments and other properties. The trajectories can be loaded and dumped in JSON, H5MD,[44] and plain text formats.

Molecules for which XYZ coordinates are provided can be transformed in several supported descriptors: inverse internuclear distances and their version normalized relative to the equilibrium structure (RE)[26], Coulomb matrix,[45, 46] and their variants.[29]

MLatom also has separate statistics routines to calculate different error measures and perform other data analyses.[29] Routines for preparing common types of plots, such as scatter plots and spectra, are available too.

## 4 Models and methods

Any of the simulations needs a model that provides the required output for a given input. The architecture and algorithms behind the models can be designed by an expert or chosen from the available selection. ML models typically require training to find their parameters before they can be used for simulations. Some of these models, such as universal MLPs of ANI family,[2, 33-35] are already pre-trained for the user who does not have to train them. This is similar to QM methods, commonly used out-of-the-box without tuning their parameters. In MLatom, we call a *method* any model that can be used out-of-the-box for simulations. Both pre-trained ML models and QM methods belong to the methods in MLatom's terminology, which is reflected in the keyword names. This model type also includes hybrid pre-trained ML and QM methods. Below, we overview models available in MLatom when writing this article, the selection of available methods and models with provided architectures that need to be trained, and the ways to design custom models (Figure 4).

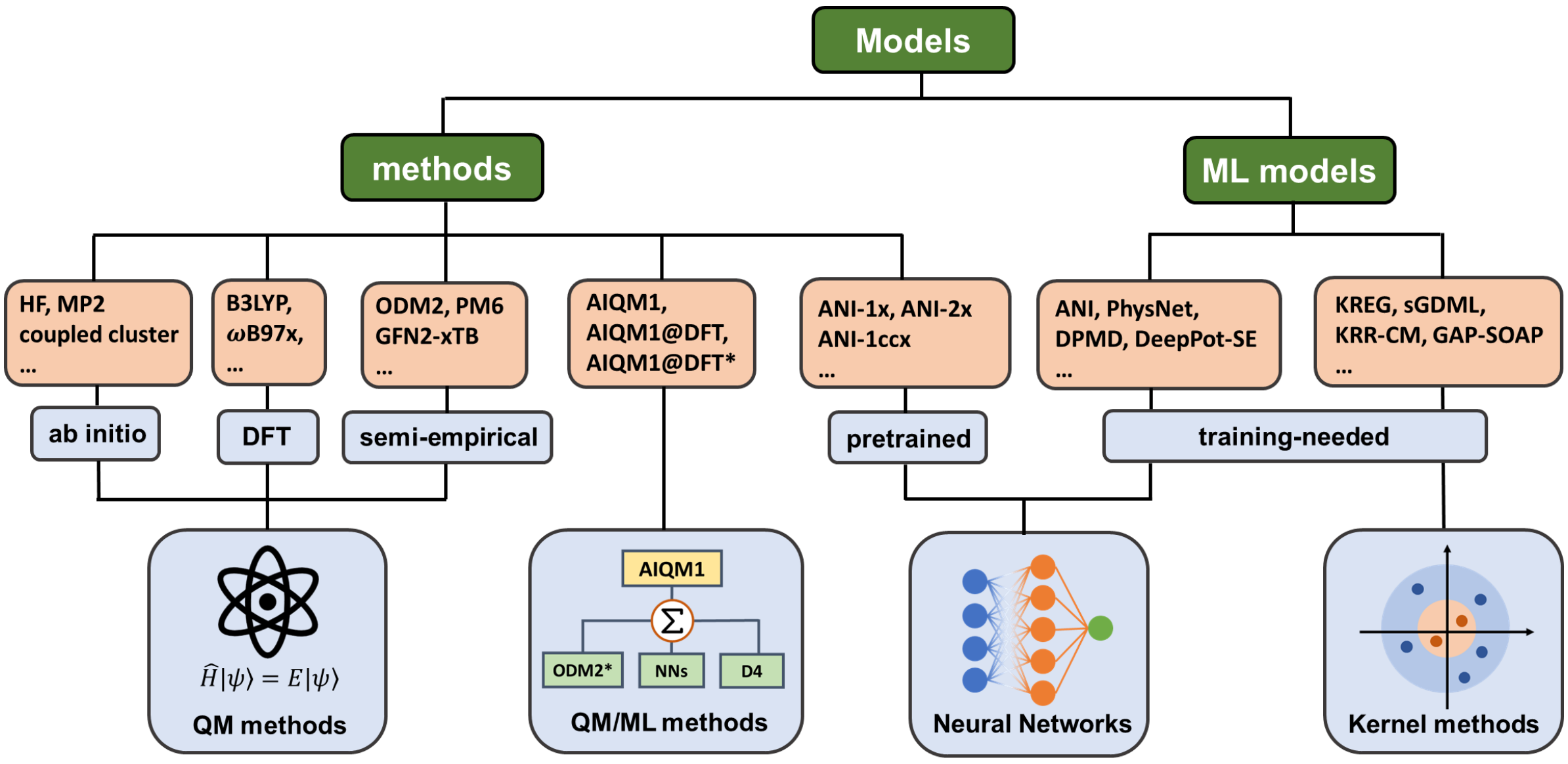

**Figure 4.** Overview of different model types in MLatom.

## *4.1 Methods*

MLatom provides access to a broad range of methods through interfaces to many third-party state-of-the-art software packages:

- Pre-trained ML models:
  - Universal potentials ANI-1ccx,[34] ANI-1x,[33] ANI-2x,[35] ANI-1x-D4, and ANI-2x-D4. ANI-1ccx is the most accurate and approaches gold-standard CCSD(T) accuracy. We have seen an example of its use in geometry optimization in Figure 2. Other methods approach the density functional theory (DFT) level. ANI-1ccx and ANI-1x are limited to CHNO elements, while ANI-2x can be used for CHNOFClS elements. We allow the user to use D4-dispersion-corrected universal ANI potentials that might be useful for noncovalent complexes. D4 correction[47] is taken for ωB97X functional[48] used to generate data for pre-training ANI-1x and ANI-2x. ANI models are provided via an interface to TorchANI[2] and D4 corrections via the interface to dftd4[49]. These methods are limited to predicting energies and forces for neutral closed-shell compounds in their ground state. MLatom reports uncertainties for calculations with these methods based on the standard deviation between neural network (NN) predictions.[36]

- Special ML-TPA model for predicting the two-photon absorption (TPA) cross sections.[31]
- Hybrid QM/ML methods: AIQM1, AIQM1@DFT, and AIQM1@DFT*.[32] More transferable and accurate than pre-trained ML models but slower (the speed of semi-empirical QM methods which are still much faster than DFT). AIQM1 is approaching gold-standard CCSD(T) accuracy, while AIQM1@DFT and AIQM1@DFT* target the DFT accuracy for neutral, closed-shell molecules in their ground state. All these methods are limited to the CHNO elements. AIQM1 and AIQM1@DFT include explicit D4-dispersion corrections for ωB97X functional while AIQM1@DFT* does not. They also include modified ANI-type networks and modified semi-empirical QM method ODM2[50] (ODM2*, provided by either the MNDO[51] or Sparrow[52] program). These methods can also be used to calculate charged species, radicals, excited states, and other QM properties such as dipole moments, charges, oscillator strengths, and nonadiabatic couplings. MLatom reports uncertainties for calculations with these methods based on the standard deviation between NN predictions.[36]
- A range of established QM methods from *ab initio* (e.g., HF, MP2, coupled cluster, *etc*.) to DFT (e.g., B3LYP,[53, 54] ωB97X,[48] etc.) via interfaces to PySCF[55] and Gaussian[55].
- A range of semi-empirical QM methods (GFN2-xTB,[56] OM2,[57] ODM2,[50] AM1,[58] PM6,[59] *etc*.) via interfaces to the xtb,[60] MNDO,[51] and Sparrow[52] programs.
- A special composite method CCSD(T)*/CBS[34] extrapolating CCSD(T) to the complete basis set via an interface to Orca[61, 62]. This method is relatively fast and accurate. It allows the user to check the quality of calculations with other methods and generate robust reference data for ML. This method was used to generate the reference data for AIQM1 and ANI-1ccx.

### *4.2 Available standard models needing training*

The field of MLPs is very rich in models. Hence, the user can often choose one of the popular MLP architectures reported in literature rather than developing a new one. MLatom provides a toolset of MLPs from different types (see Ref. [39] for an overview and Ref. [29] for implementation details). These supported types can be categorized in a simplified scheme as

- Models based on kernel methods (KMs)[63] with global descriptors to which (p)KREG,[26, 63] sGDML,[65] and KRR-CM[45, 46] belong as well as with local descriptors represented by only GAP[66]-SOAP[67].
- Models based on neural networks (NNs) with fixed local descriptors to which ANI-type MLPs[2] and DPMD[68] belong and with learned local descriptors represented by PhysNet[69] and DeepPot-SE[70].

Any of these models can be trained and used for simulations, e.g., geometry optimizations or dynamics. MLatom also supports hyperparameter optimization with many algorithms including grid search,[22] Bayesian optimization via the hyperopt package,[71, 72] and standard optimization algorithms available in SciPy[73]. Generalization errors of the resulting models can also be evaluated in standard ways (hold-out and cross-validation). More on this in a dedicated Section 6.

### *4.3 Custom models based on kernel methods*

MLatom also provides the flexibility of training custom models based on kernel ridge regression (KRR) for a given set of input vectors **x** or XYZ coordinates and any labels **y**.[74, 75] If XYZ coordinates are provided, they can be transformed in one of the several supported descriptors (e.g., inverse internuclear distances and their version normalized relative to the equilibrium structure (RE), and Coulomb matrix). The user can choose from one of the implemented kernel functions, including the linear,[22, 75, 76] Gaussian,[22, 75, 76] exponential,[22, 75, 76] Laplacian,[22, 75, 76] and Matérn[22, 75-77] as well as periodic[76, 78, 79] and decaying periodic[76, 78, 80] functions, which are summarized in Table 1. These kernel functions $k(\mathbf{x}, \mathbf{x}_j; \mathbf{h})$ are key components required to solve the KRR problem of finding the regression coefficients $\alpha$ of the approximating function $\hat{f}(\mathbf{x}; \mathbf{h})$ of the input vector **x**:[74, 75]

$$\hat{f}(\mathbf{x}; \mathbf{h}) = \sum_{j=1}^{N_{\text{tr}}} \alpha_j k(\mathbf{x}, \mathbf{x}_j; \mathbf{h}). \quad (1)$$

The kernel function, in most cases, has hyperparameters **h** to tune, and they can be viewed as measuring similarity between the input vector **x** and all of $N_{\text{tr}}$ training points $\mathbf{x}_j$ (both vectors should be of the same length $N_x$). In addition to the hyperparameters in the kernel function, all KRR models have at least one more regularization parameter $\lambda$ used during training to improve the generalizability.

**Table 1.** Summary of the available kernel functions for solving the kernel ridge regression problem (Eq. 1) as implemented in MLatom.

| Kernel function | Formula | Hyperparameters in kernel function |
| --- | --- | --- |
| Linear | $k(\mathbf{x},\mathbf{x}_j) = \mathbf{x}^{\mathrm{T}}\mathbf{x}_j$ | |
| Gaussian | $k(\mathbf{x},\mathbf{x}_j) = \exp\left(-\frac{1}{2\sigma^2}\sum_{s}^{N_x}(x_s - x_{j,s})^2\right)$ | $\sigma > 0$, length scale |
| exponential | $k(\mathbf{x},\mathbf{x}_j) = \exp\left(-\frac{1}{\sigma}\left[\sum_{s}^{N_x}(x_s - x_{j,s})^2\right]^{1/2}\right)$ | $\sigma > 0$, length scale |
| Laplacian | $k(\mathbf{x},\mathbf{x}_j) = \exp\left(-\frac{1}{\sigma}\sum_{s}^{N_x}\lvert x_s - x_{j,s}\rvert\right)$ | $\sigma > 0$, length scale |
| Matérn | $k(\mathbf{x},\mathbf{x}_j) = \exp\left(-\frac{1}{\sigma}\left[\sum_{s}^{N_x}(x_s - x_{j,s})^2\right]^{\frac{1}{2}}\right) \times \sum_{k=0}^{n}\frac{(n+k)!}{(2n)!}\binom{n}{k} \times \left(\frac{2}{\sigma}\left[\sum_{s}^{N_x}(x_s - x_{j,s})^2\right]^{1/2}\right)^{n-k}$ | $\sigma > 0$, length scale; $n$ is a non-negative integer |
| periodic | $k(\mathbf{x},\mathbf{x}_j) = \exp\left(-\frac{2}{\sigma^2}\sin^2\left\{\frac{\pi}{p}\left[\sum_{s}^{N_x}(x_s - x_{j,s})^2\right]^{1/2}\right\}\right)$ | $\sigma > 0$, length scale; $p > 0$, period |
| decaying periodic | $k(\mathbf{x},\mathbf{x}_j) = \exp\left(-\frac{1}{2\sigma^2}\sum_{s}^{N_x}(x_s - x_{j,s})^2 - \frac{2}{\sigma_p^2}\sin^2\left\{\frac{\pi}{p}\left[\sum_{s}^{N_x}(x_s - x_{j,s})^2\right]^{1/2}\right\}\right)$ | $\sigma > 0$, length scale; $p > 0$, period; $\sigma_p > 0$, length scale for the periodic term |

### *4.4 Composite models*

Often, it is beneficial to combine several models. One example of such composite models is based on Δ-learning[23] where the low-level QM method is used as a baseline which is corrected by an ML model to approach the accuracy of the target higher-level QM method. Another example is ensemble learning[81] where multiple ML models are created, and their predictions are averaged during the simulations to obtain more robust results and use in the query-by-committee strategy of active learning.[82] Both of these concepts can also be combined in more complex workflows as exemplified by the AIQM1 method[32] which uses the NN ensemble as a correcting Δ-learning model and the semi-empirical QM method as the baseline. To easily implement these workflows, MLatom allows the construction of the composite models as model trees; see an example for AIQM1 in Figure 5.

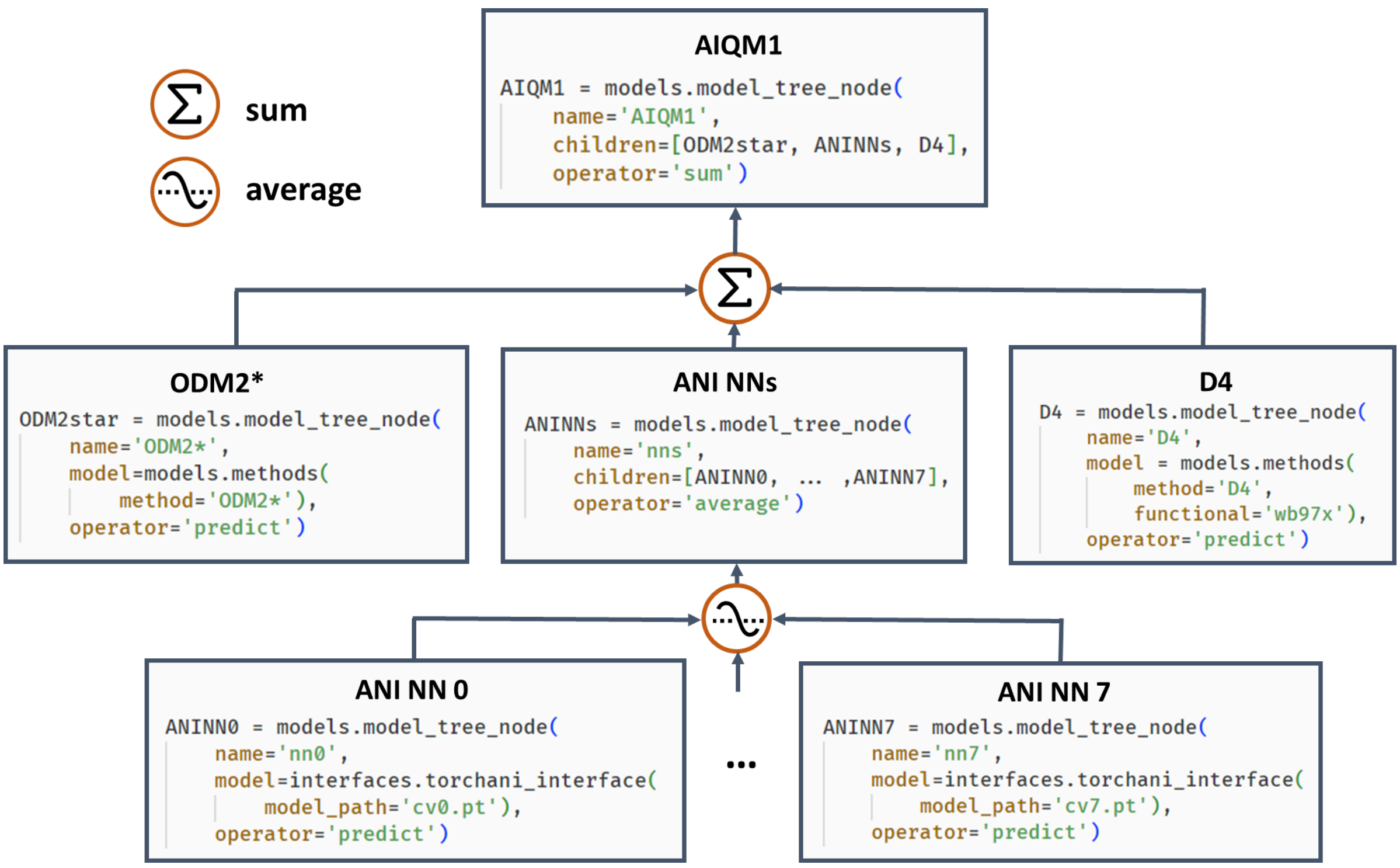

**Figure 5.** Composite models can be constructed as a model tree in MLatom. Here an example is shown for the AIQM1 method where the root parent node comprises 3 children, the semi-empirical QM method ODM2*, the NN ensemble, and additional D4 dispersion correction. NN ensemble in turn is a parent of 8 ANI-type NN children. Predictions of parents are obtained by applying an operation 'average' or 'sum' to children's predictions. The code snippets are shown, too.

Other examples of possible composite models are hierarchical ML,[25] which combines several (correcting) ML models trained on (differences between) QM levels, and self-correction,[26] when each next ML model corrects the prediction by the previous model.

## 5 Simulations

MLatom supports a range of simulation tasks such as single-point simulations, geometry optimizations, frequency and thermochemistry calculations, molecular and quantum dynamics, one- and two-photon absorption and (ro)vibrational spectra simulations (Figure 6). Most of them need any model that can provide energies and energy derivatives (gradients and Hessians).

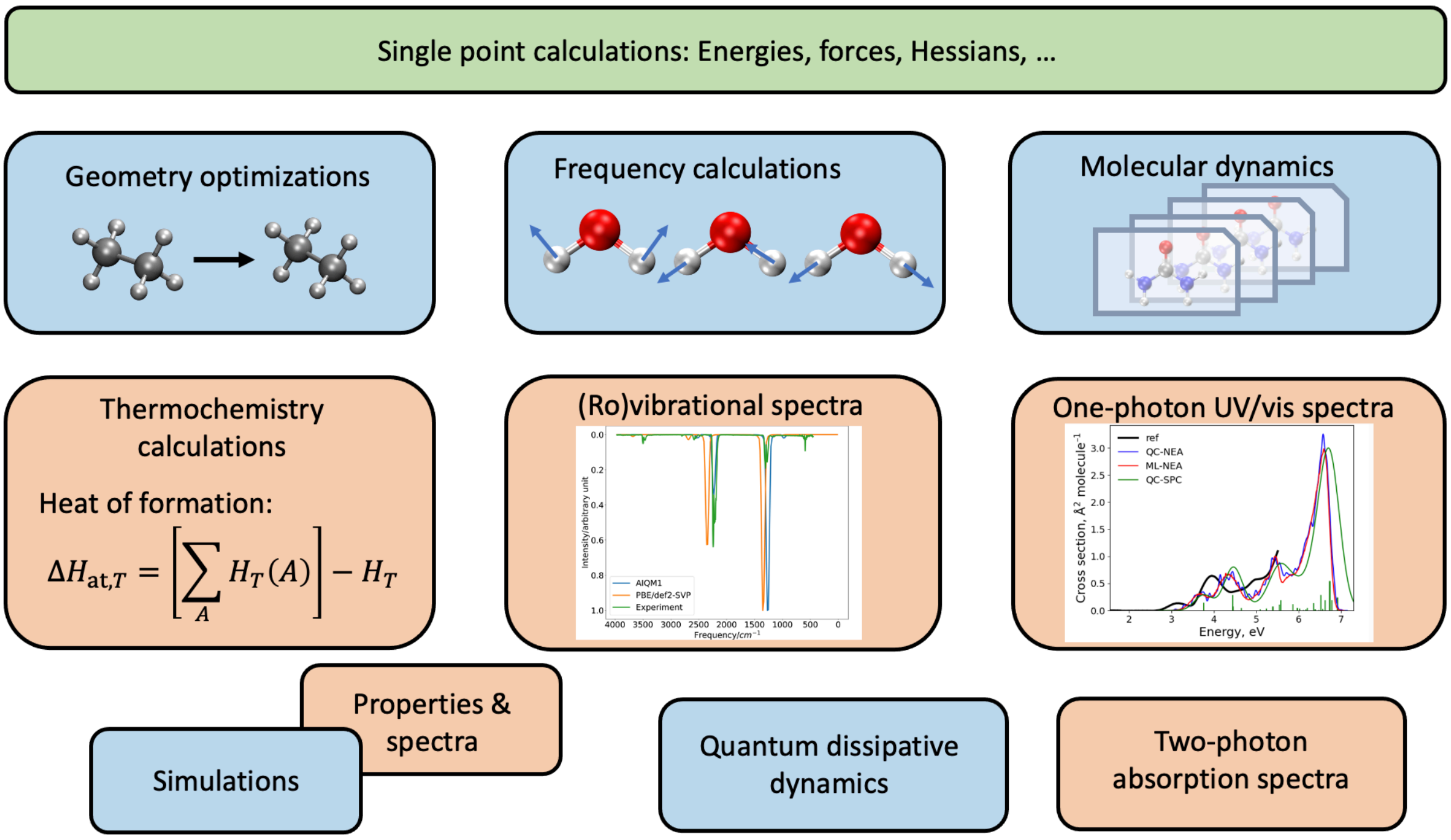

**Figure 6.** Overview of simulation tasks in MLatom. The inset in one-photon UV/vis spectra is reproduced from Ref. 29 under the CC-BY-4.0 license.

### *5.1 Single-point calculations*

Single-point calculations are calculations of quantum mechanical properties — mostly energies and energy gradients, but also Hessians, charges, dipole moments, *etc.* — for a single geometry. These calculations are very common in ML research in computational chemistry as they are used both to generate the reference data with QM methods for training and validating ML and to make inferences with ML to validate the trained model and generate required data for new geometries. MLatom is a convenient tool to perform single-point calculations not just for a single geometry, as in many QM packages, but for data sets with many geometries.

### *5.2 Geometry optimizations*

Locating stationary points on PES, such as energy minima and transition states, is crucial for understanding the molecular structure and reactivity. Hence, geometry optimizations are among the most important and frequent tasks in computational chemistry. MLatom can locate energy minima and transition states (TS) with any models providing energies and gradients. An example of geometry optimization is given in Figure 2. Hessians are also required for the Berny TS optimization algorithm. Once the TS is located, the user can follow the intrinsic reaction coordinate (IRC)[38] to check its nature. Geometry optimizations can be performed with many algorithms provided by the interfaces to SciPy,[73] ASE,[18] or Gaussian[55]. TS search can be performed with the dimer method[83] in ASE and the Berny algorithm[84] in Gaussian. IRC calculations can only be performed with the interface to Gaussian.

The seamless integration of the variety of QM and ML methods for performing geometry optimizations is advantageous because it allows the use of methods from interfaced programs that do not implement some of such simulation tasks by themselves. For example, MLatom can be used to perform TS search with the GFN2-xTB method via an interface to the xtb program, while there is no option for TS search with the latter program. Similarly, Sparrow, which provides access to many semi-empirical methods, can only be used for single-point calculations. Since analytical gradients and Hessians are not available for many models and implementations, MLatom also implements a finite-difference numerical differentiation, further expanding the applicability of the models for geometry optimizations.

### *5.3 Frequency calculations*

Simulation of vibrational frequencies is another common and important task in computational chemistry as it is useful to additionally verify the nature of stationary points, visualize molecular vibrations, calculate zero-point vibrational energy (ZPE) and thermochemical properties, as well as obtain spectroscopic information, which can be compared to experimental vibrational spectra. These calculations can be performed within the ridge-rotor harmonic approximation via an adapted TorchANI implementation[2] and Gaussian[55] interface. The latter also allows the calculation of anharmonic frequencies using the second-order perturbative approach.[85]

Similarly to geometry optimizations, MLatom can perform these simulations with any model — ML and QM or their combination — that provides energies. Calculations also need

Hessian, and wherever available, analytical Hessian is used. If it is unavailable, semi-analytical (with analytical gradients) or fully numerical Hessian can be calculated.

### *5.4 Thermochemistry calculations*

Thermochemical properties such as enthalpies, entropies, and Gibbs free energies can be derived from frequency calculations. In turn, enthalpies can be used to calculate heats (enthalpies) of formation. MLatom uses the scheme analogous to those employed in the *ab initio*[86] and semi-empirical QM calculations[50] to derive heats of formation:

$$\Delta H_{\mathrm{f},T} = \left[\sum_{A} \Delta H_{\mathrm{f},T}(A)\right] - \Delta H_{\mathrm{at},T}, \tag{2}$$

where $\Delta H_{\mathrm{f},T}(A)$ is the experimental enthalpies of formation of free atom A, and $\Delta \mathrm{H}_{\mathrm{at},T}$ is the atomization enthalpy. In AIQM1 and ANI-1ccx, we use the same $\Delta H_{\mathrm{f},T}(A)$ values as other semi-empirical QM methods, i.e., 52.102, 170.89, 113.00, 59.559 kcal/mol for elements H, C, N, O, respectively.[51]

The atomization enthalpy $\Delta H_{\mathrm{at},T}$ can be obtained from the difference between molecular $H_T$ and atomic absolute enthalpies $H_T(A)$:

$$\Delta H_{\mathrm{at},T} = \left[\sum_{A} H_T(A)\right] - H_T. \tag{3}$$

Analogous to *ab initio* methods, harmonic-oscillator and rigid-rotor approximations are explicitly considered in the calculation of absolute enthalpies:

$$H_T = E_{tot} + ZPVE + E_{\mathrm{trans},T} + E_{\mathrm{rot},T} + E_{\mathrm{vib},T} + RT, \tag{4}$$

$$H_T(A) = E(A) + E_{\mathrm{trans},T}(A) + RT, \tag{5}$$

where $E_{\mathrm{tot}}$ and $E(A)$ are the total energy of the molecule and free atom, respectively, and ZPVE is the zero-point vibrational energy. $E_{\mathrm{trans},T}$, $E_{\mathrm{rot},T}$ and $E_{\mathrm{vib},T}$ are the translational, rotational, and vibrational thermal contributions, and *R* is the gas constant.

The scheme requires the knowledge of free atom energies $E(A)$. Any model able to calculate them can be used for predicting heats of formation. This is straightforward for QM methods but not for ML-based models that are usually trained on molecular species. We have previously fitted free atom energies (see Table 2) for AIQM1 and ANI-1ccx methods to the experimental data set.[32, 36] As a result, both methods can provide heats of formation close to chemical accuracy with speed orders of magnitude higher than that of alternative high-accuracy QM methods. In addition, we provide an uncertainty quantification scheme based on the deviation of NN predictions in these methods to tell the users when the predictions are confident. This was useful to find errors in the experimental data set of heats of formation.[36]

**Table 2.** The atomic energies (in Hartree) of AIQM1 and ANI-1ccx used in heats of formation calculations.[32, 36]

| Element | AIQM1 | ANI-1ccx |
|---|---|---|
| H | -0.50088038 | -0.50088088 |
| C | -37.79221710 | -37.79199048 |
| N | -54.53360298 | -54.53379230 |
| O | -75.00986203 | -75.00968205 |

An example of using MLatom to calculate heats of formation with the AIQM1 and B3LYP/6-31G* methods is shown in Figure 7. AIQM1 is both faster and more accurate than B3LYP, as can be seen by comparing the values with the experiment. This is also consistent with our previous benchmark.[36]

**Input**

sp.xyz

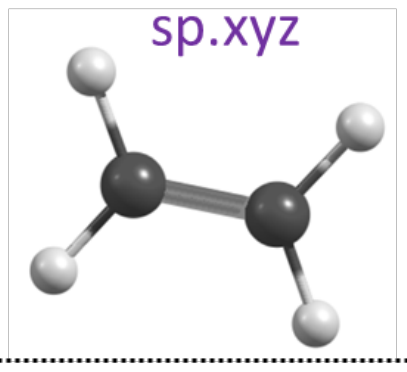

```
6

C      0.00000    0.00000    0.00000
C      1.33900    0.00000    0.00000
H      1.90372    0.00000    0.92880
H     -0.56472    0.00000    0.92880
H     -0.56472    0.00000   -0.92880
H      1.90372    0.00000   -0.92880
```

**Command-line input file**

```
method=AIQM1         # define AIQM1 method
qmprog=mndo          # define QM program for method
freq                 # request frequency calculation
xyzfile=sp.xyz       # file with input geometry

method=B3LYP/6-31G*  # define B3LYP/6-31G* method
qmprog=pyscf         # define QM program for method
freq                 # request frequency calculation
xyzfile=sp.xyz       # file with input geometry
```

**Python API**

```
import mlatom as ml

aiqm1 = ml.models.methods(method='AIQM1', qm_program='MNDO')
mol = ml.data.molecule.from_xyz_file('sp.xyz')
ml.simulations.thermochemistry(model=aiqm1, molecule=mol)
print(mol.DeltaHf298)

b3lyp = ml.models.methods(method='B3LYP/6-31G*', program='pyscf')
mol = ml.data.molecule.from_xyz_file('sp.xyz')
ml.simulations.thermochemistry(model=b3lyp, molecule=mol)
print(mol.DeltaHf298)
```

**Output**

| Method | $\Delta H_f^{experiment}$ (kcal/mol) | $\Delta H_f^{calculated}$ (kcal/mol) | Time (s) |
|---|---|---|---|
| AIQM1 | 12.50 | 12.47 | 3.44 |
| B3LYP/6-31G* | 12.50 | 14.12 | 11.26 |

**Figure 7.** Calculation of heats of formation of ethylene with AIQM1 and B3LYP/6-31G* (from the interface to PySCF) compared to the experiment[87].

### *5.5 Molecular dynamics*

Molecular dynamics propagates nuclear motion based on the equation of motion according to the classical mechanics.[88] This requires the knowledge of forces acting on nuclei, which are typically derived as the negative of the potential energy gradients (i.e., negative of the derivatives of the model for potential energies) for conservative forces. Due to the high cost of the approach, it is most commonly used with molecular mechanics force fields,[89] but often, calculations based on QM methods are possible in variants called *ab initio* or Born–Oppenheimer MD (BOMD).[88] The proliferation of ML potentials makes it possible to perform BOMD quality dynamics at a cost comparable to molecular mechanics force fields or much faster than commonly used DFT-based BOMD.[39-43] For example, the AIQM1 method is faster than DFT and the IR spectra obtained from AIQM1 MD are of higher quality (Figure 8).[90]

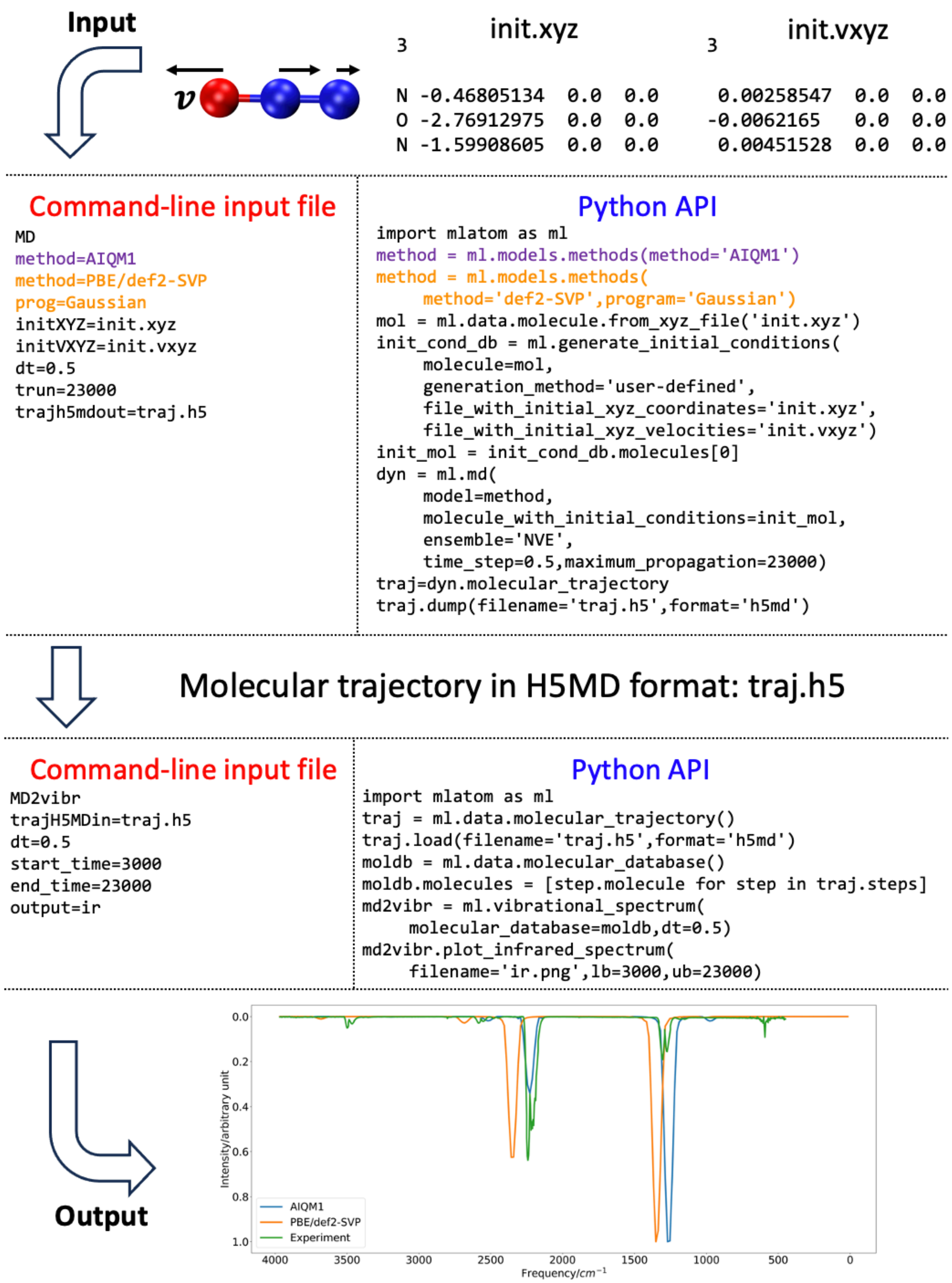

**Figure 8.** Propagation of MD with AIQM1 and PBE/def2-SVP (from interface to Gaussian) and the IR spectra of $N_2O$ molecule derived from trajectories. MLatom generates spectra for each method, here the results are collated and shown together with the experimental spectrum[91] for comparison.

MLatom has a native implementation of MD supporting any kind of model that provides forces, not necessarily conservative.[90] Currently, simulations in NVE and NVT ensembles[92], based on the velocity Verlet algorithm,[93] are possible. NVT simulations can be carried out with the Andersen[92, 94] and Nosé–Hoover[95, 96] thermostats. Trajectories can be saved in different formats, including plain text, JSON and, more compact H5MD[29] database formats. The Nosé–Hoover thermostat is a deterministic thermostat that couples the system to a thermal bath through extra terms in the Hamiltonian. Its theory and implementation details are described elsewhere.[90] Here, we briefly mention the relevant methodology[92, 94] used in the Andersen thermostat. In this thermostat, the system is coupled to a heat bath by stochastically changing the velocity of each atom. The changing frequency (or collision frequency) is controlled by a tunable parameter $v$. The collisions follow the Poisson distribution so that the probability of changing the velocity of each atom during a time step $\Delta t$ is $v\Delta t$. If the atoms collide, new velocities will be assigned to them, sampled from a Maxwell–Boltzmann distribution at target temperature $T$.

MD trajectories can be propagated in parallel, dramatically speeding up the calculations. In addition, we made an effort to better integrate the KREG model implemented in Fortran into the main Python-based MLatom code which makes MD with KREG very efficient.

Note that MD can also be propagated without forces using the concept of the 4D-spacetime AI atomistic models, which directly predict nuclear configurations as a function of time.[79] Our realization of this concept, called the GICnet model, is currently available in a publicly available development version of MLatom version.[79]

The above implementations can propagate MD on an adiabatic potential energy surface, i.e., typically, for ground-state dynamics. Nonadiabatic MD based on the trajectory surface hopping algorithms can also be performed with the help of MLatom, currently, via Newton-X[96]'s interface to MLatom.[27, 97, 98] MLatom also supports quantum dissipative dynamics as described in the next Section 5.6.

### *5.6 Quantum dissipative dynamics*

It is often necessary and beneficial to treat the entire system quantum mechanically and also include the environmental effects.[100] This is possible via many quantum dissipative dynamics (QD) algorithms, and an increasing number of ML techniques were suggested to accelerate such simulations.[98] MLatom allows performing several unique ML-accelerated QD simulations using either a recursive scheme based on KRR[101] or a conceptually different AI-QD approach[102] predicting the trajectories as a function of time or OSTL technique[103] outputting the entire trajectories in one shot. These approaches are enabled via an interface to a specialized program MLQD.[104]

In the recursive KRR scheme, a KRR model is trained, establishing a map between future and past dynamics. This KRR model, when provided with a brief snapshot of the current dynamics, can be leveraged to forecast future dynamics. In the AIQD approach, a convolution neural network (CNN) model is trained mapping simulation parameters and time to the corresponding system's state. Using the trained CNN model, the state of the system can be predicted at any time without the need to explicitly simulate the dynamics. Similarly, the ultra-fast OSTL method utilizes CNN-based architecture and, based on simulation parameters, predicts future dynamics of the system's state up to a predefined time in a single shot. In addition, as optimization is a key component in training, users can optimize both KRR and CNN models using MLatom's grid search functionality for KRR and Bayesian optimization via the hyperopt[71] library for CNN. Moreover, we also incorporate the auto-plotting functionality, where the predicted dynamics is plotted against the provided reference trajectory.

### *5.7 Rovibrational (infrared and power) spectra*

Rovibrational spectra can be calculated in several ways with MLatom. The simplest one is by performing frequency calculations on an optimized molecular geometry. This requires any model providing Hessians and, preferably, dipole moments. Another one is performing molecular dynamics simulations with any model providing energy gradients and, then, post-processing the trajectories.

Both frequency calculations and the MD-based approach require the model to also provide dipole moments to calculate absorption intensities. If no dipole moments are provided, only frequencies are available, or, in the case of MD, only power spectra rather than IR can be obtained. The IR spectra are obtained via the fast Fourier transform using the autocorrelation function of dipole moment[104, 105] with our own implementation.[90] The power spectra only need the fast Fourier transform,[105] which is also implemented[79] in MLatom.

We have previously shown[90] that the high quality of the AIQM1 method results in rather accurate IR spectra obtained from MD simulations compared to spectra obtained with a representative DFT (which is also substantially slower; see example in Figure 8) or a semi-empirical QM method.

### *5.8 One-photon UV/vis absorption spectra*

UV/vis absorption spectra simulations are computationally intensive because they require calculating excited-state properties. In addition, better-quality spectra can be obtained via the nuclear ensemble approach (NEA),[107] which necessitates the calculation of excited-state properties for thousands of geometries for high precision. MLatom implements an interpolation ML-NEA scheme[30] that improves the precision of the spectra with a fraction of the computational cost of traditional NEA simulations. Currently, the ML-NEA calculations are based on interfaces to Newton-X[96] and Gaussian[55] and utilize the sampling of geometries from a harmonic Wigner distribution[108]. This scheme also automatically determines the optimal number of required reference calculations, providing a user-friendly, black-box implementation of the algorithm.[29]

### *5.9 Two-photon absorption*

Beyond one-photon absorption, MLatom has an implementation of a unique ML approach for calculating two-photon absorption (TPA) cross sections of molecules just based on their SMILES strings,[44] which are converted into the required descriptors using the interface to RDKit,[109] and solvent information.[31] This ML-TPA approach is very fast with accuracy comparable to much more computationally intensive QM methods. We provide a ML model pre-trained on experimental data. ML-TPA was tested in real laboratory settings and shown to provide a good estimate for new molecules not present in the training experimental database.

## 6 Machine learning

In Sections 4 and 5, we discussed the supported types of models and how they can be applied to simulations. Here, we briefly overview the general considerations for training and validating the ML models with MLatom. The models share the standard MLatom's conventions for input, output, training, hyperparameter optimization, and testing, which allows to conveniently switch from one to another model and benchmark them.

### *6.1 Training*

To create an ML model, the user has to choose and train the ML model and prepare data. MLatom provides many tools for different stages of this process. The model can be either chosen from a selection of provided types of ML models with pre-defined architecture or customized based on available algorithms and preset models. Once a model is chosen, it must be trained, and, in many cases, it is advisable or even required (particularly in the case of the kernel methods) to optimize its hyperparameters, which can be done as explained in Section 6.2.

For training, the data set should be appropriately prepared. MLatom has strict naming conventions for data set splits to avoid any confusion when changing and comparing different model types. All the data that is used directly or indirectly for creating a ML model is called the training set. This means that the validation set, which can be used for hyperparameter optimization or early stopping during NN training, is a subset of the training set. Thus, the part of the training set remaining after excluding the validation set is called the sub-training set and is actually used for training the model, i.e., optimizing model parameters (weights in NN terminology and regression coefficients in kernel methods terminology).

MLatom can split the training data set into the sub-training and validation data subsets or create a collection of these subsets via cross-validation.[24, 29] The sampling into the subsets can be performed randomly or using furthest-point or structure-based sampling.

In the case of kernel methods, the final model in MLatom is typically trained on the entire training set after the hyperparameter optimization. This is possible because the kernel methods have a closed, analytical solution to finding their regression coefficients, and after hyperparameters are appropriately chosen, overfitting can be mitigated to a great extent. In the case of NNs, the final model is the one trained on the sub-training set because it would be too dangerous to train on the entire training set without any validation subset to check for the signs of overfitting.

### *6.1.1 Training pre-defined types of ML models*

Most pre-defined types of ML models, such as ANI-type or KREG models, expect XYZ molecular coordinates as input. This should be either provided by the user or can be obtained using MLatom's conversion routines, e.g., from the SMILES strings,[110] which rely on OpenBabel[111]'s Pybel API. These models have a default set of hyperparameters, but especially in the case of kernel methods such as KREG, it is still strongly advised to optimize them. The models can be, in principle, trained on any molecular property. Most often, they are used to learn PESs and, hence, require energy labels in the training set. The PES model accuracy can be greatly improved if the energy gradients are also provided for training. Thus, the increased training time is usually justified.[39, 112] An example of training and testing the KREG and DPMD models on a data set with energies and energy gradients for the urea molecule in the WS22 database[113] is shown in Figure 9. The KREG model is both faster to train and more accurate, which is a typical situation for small-size molecular databases, while for larger databases, NN-based models might be preferable.[39]

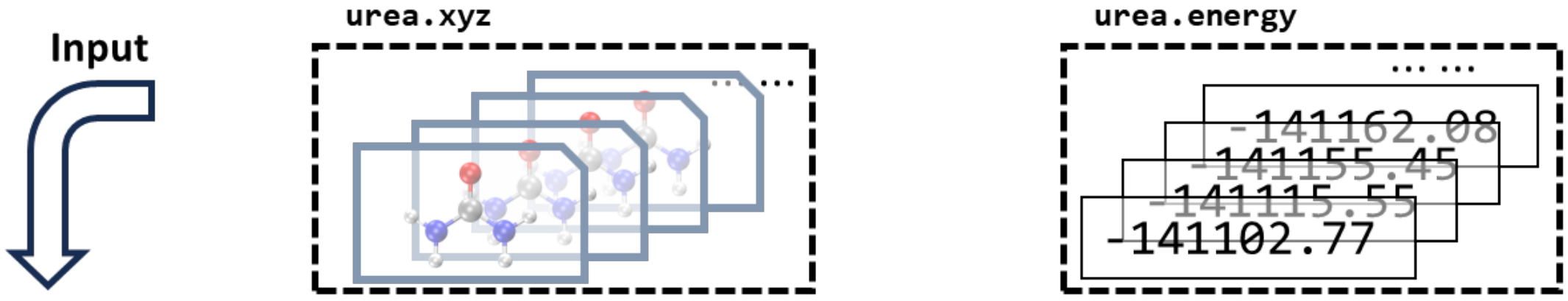

Command-line input file

```
# KREG model
estAccMLmodel
XYZfile=urea.xyz
Yfile=urea.energy
MLmodelType=KREG
sigma=opt
lambda=opt

# DeepPot-SE model
estAccMLmodel
XYZfile=urea.xyz
Yfile=urea.energy
MLmodelType=DeepPot-SE
```

Python API

```python
import mlatom as ml

# prepare data
molDB = ml.data.molecular_database.from_xyz_file("urea.xyz")  # load geometry data
molDB.add_scalar_properties_from_file("urea.energy", "energy")  # load energy data
subtraining_molDB, validation_molDB, test_molDB = ml.data.sample(
    number_of_splits=3,  # split database for subtraining, validataion, and test
    fraction_of_points_in_splits=[0.64, 0.16, 0.2],  # split ratio
    molecular_database_to_split=molDB,
    sampling="random",
)
# define a KREG model
model = ml.models.kreg()
# optimize hyperparameters
model.optimize_hyperparameters(
    subtraining_molecular_database=subtraining_molDB,
    validation_molecular_database=validation_molDB,
    optimization_algorithm="grid",
    hyperparameters=["lambda", "sigma"],
    training_kwargs={"property_to_learn": "energy"},
)
# or define a DeepPot-SE model
model = ml.models.dpmd()
# train the final model
kreg_model.train(
    molecular_database=subtraining_molDB + validation_molDB,
    property_to_learn="energy"
)
# make predictions on the test set
kreg_model.predict(test_molDB, property_to_predict="estimated_energy")
# analyze the RMS error
ml.stats.rmse(
    test_molDB.get_properties("energy"), test_molDB.get_properties(" estimated _energy")
)
```

Output

| Model type | RMSE (kcal/mol) | Time (s) |
|---|---|---|
| KREG | 0.835 | 2.77 |
| DeepPot-SE | 5.498 | 43136.35 |

**Figure 9.** Side-by-side comparison of the usage of MLatom in both command-line mode and via Python API for training and testing the KREG and DeepPot-SE models on a 1000-point data set on the urea molecular PES data set randomly sampled from the WS22 database. Hyperparameter optimization of the KREG model is required is also shown. Calculations were run on 36 Intel(R) Xeon(R) Gold 6240 CPU @ 2.60GHz.

### 6.1.2 Designing and training custom ML models

MLatom's user can also create models on any set of input vectors and labels using a variety of KRR kernel functions. In this case, hyperparameter optimization is strongly advised too. In all other aspects, training of such KRR models is similar to training the pre-defined models, i.e., the preparation of the data set is also performed by splitting it into the required subsets for training and validation.

Importantly, the user can construct models of varying complexity by using a model tree implementation. Special cases of such composite models are Δ-learning and self-correcting models and they can be trained similarly to other ML models by supplying input vectors or XYZ coordinates and labels. In the case of Δ-learning, the user needs to supply the baseline values. For other, more complicated models, the user must train and combine each component separately.

## 6.2 Hyperparameter optimization

The performance of ML models strongly depends on the chosen hyperparameters such as the regularization parameters and number of layers in NNs. Hence, it is often necessary to optimize the hyperparameters to achieve reasonable results and to improve the accuracy. The hyperparameter optimization commonly requires multiple trainings, making it an expensive endeavor, and caution must paid in balancing performance/cost issues.

MLatom can optimize hyperparameters by minimizing the validation loss using one of the many available algorithms. The validation loss is usually based on the error in the validation set which can be a single hold-out validation set, or a combined cross-validation error.

For few hyperparameters, the robust grid search on the log or linear scale can be used to find optimal values. It is a common choice for kernel methods (see Figure 9 for an example of optimizing hyperparameters of the KREG model which is the kernel method). For a larger number of hyperparameters, other algorithms are recommended instead. Popular choices are Bayesian optimization with the Tree-structured Parzen Estimator (TPE)[72] and many SciPy optimizers.

The choice of the validation loss also matters. In most cases, MLatom minimizes the root-mean-squared error (RMSE) for the labeled data. However, when multiple labels are provided, i.e., energies and energy gradients for learning PES, the choice should be made on how to combine them in the validation loss. By default, MLatom calculates the geometric mean of the RMSEs for energies and gradients.[29] The users can also choose a weighted sum of RMSEs, but in this case, they must choose the weight. In addition, the user can supply MLatom with any custom validation loss function, which can be arbitrarily complicated.

### *6.3 Evaluating models*

Once the model has been trained, it is common to evaluate its generalization ability before deploying it in production simulations. MLatom provides dedicated options for such evaluations. The simplest and one of the most widespread approaches is calculating the error for the independent hold-out test set not used in the training. To emphasize, in MLatom terminology, the test set has no overlap with the training set, which might consist of the sub-training and validation subsets.[29] Alternatively, cross-validation and its variant leave-one-out cross-validation are recommended whenever computationally affordable, especially for small data sets. MLatom provides a broad range of error measures for the test set, including RMSE, mean absolute error (MAE), mean signed error, the Pearson correlation coefficient, the $R^2$ value, outliers, *etc*.[29] The testing can be performed with the training and hyperparameter optimization for most models, including Δ-learning and self-correcting models.

Since the errors depend on the size of the training set, the learning curves showing this dependence are very useful for comparing different models.[29] MLatom can generate the learning curves, which have been instrumental in preparing guidelines for choosing the ML interatomic potential.[39]

## Summary

MLatom 3 is a unique software package combining machine learning and quantum mechanical models for accelerating and improving the accuracy of computational chemistry simulations. It can be used as a black-box package accepting input files with a simple structure or as a transparent Python module enabling custom workflows. MLatom provides access to pre-trained models such as AIQM1 and ANI-1ccx aiming at high accuracy of coupled-cluster level, making them more accurate and much faster than common DFT approaches for ground-state properties of closed-shell organic molecules. Another special pre-trained model can be used to simulate two-photon absorption spectra.

The user of MLatom has an option to create their own models. Pre-defined ML architectures of the ANI-type, KREG, PhysNet, GAP-SOAP, DPMD, or sGDML make it easier. Alternatively, the custom models of varying complexity and based on combinations of both ML and QM models, such as Δ-learning can be easily built with the package. MLatom provides a toolset for training, hyperparameter optimization, and performance analysis of the models.

This wide variety of models can be used for single-point calculations on large data sets, geometry optimizations, calculation of rovibrational (frequencies, IR spectra) and thermochemical (enthalpies, entropies, heats of formation) properties, molecular dynamics, and UV/vis absorption spectra. The ML models can also be trained and used for quantum dissipative dynamics simulations.

The richness of MLatom functionality is available open source and can be exploited on the XACS cloud computing service. The package is accompanied by extensive and detailed manuals and tutorials that are developed and improved in close connection with teaching computational chemistry and machine learning in regular workshops and university courses.

**Data availability**

No data was generated for this article.

**Code availability**

The MLatom code is open-source and available both on GitHub (https://github.com/dralgroup/mlatom) and PyPI (i.e., it can be installed via the command `pip install mlatom`). The simulations can also be run on MLatom@XACS cloud computing service on https://XACScloud.com.

**Author contributions**

P.O.D. is the lead designer, developer, and maintainer of MLatom. F.G. is co-maintaining the MLatom package, implemented interfaces to third-party machine learning packages (PhysNet, DeePMD-kit, TorchANI, and GAP-SOAP), hyperopt, wrote the code for learning curve, and made numerous other improvements in MLatom. Y.F.H. co-implemented the KREG model, implemented molecular dynamics and vibrational spectra simulations, and improved many other parts of the code such as interfaces. P.Z. implemented AIQM1 and the ANI family of models (ANI-1ccx, ANI-2x, ANI-1x and their dispersion-corrected variants) through interfaces to third-party packages (MNDO, TorchANI, Sparrow) as well as geometry optimizations, frequency and thermochemistry simulations via interfaces to Gaussian, ASE, and TorchANI. Y.X.X.C. implemented interfaces to PySCF and Orca and extended thermochemical calculations to many methods. M.B. contributed to planning the implementation of MLPs and the methodology behind the ML-NEA approach. O.I. contributed to the research involving AIQM1 methods and ANI universal potentials. C.W. led the development of the ML-TPA methodology. B.X.X. implemented the ML-NEA approach and initial argument parsing routines. M.P.J. helped implement of the interfaces to TorchANI, PhysNet, DeePMD-kit, and Newton-X. Y.S., Y.D., and Y.T.C. implemented ML-TPA approach. L.Z. implemented routines for nonadiabatic dynamics and extensions of the MNDO interface to excited-state properties. S.Z. contributed to atomic properties collection and implemented some of the NN-based approaches. A.U. interfaced MLQD to MLatom. Q.Z. contributed to the program documentation and tests. Y.O. contributed to plotting routines. P.O.D. wrote the original manuscript and all authors revised and commented on the manuscript. F.G., Y.F.H., Y.X.X.C., and P.O.D. prepared the figures.

**Acknowledgments**

P.O.D. acknowledges funding by the National Natural Science Foundation of China (No. 22003051 and funding via the Outstanding Youth Scholars (Overseas, 2021) project), the Fundamental Research Funds for the Central Universities (No. 20720210092), and via the Lab project of the State Key Laboratory of Physical Chemistry of Solid Surfaces. This project is supported by Science and Technology Projects of Innovation Laboratory for Sciences and Technologies of Energy Materials of Fujian Province (IKKEM) (No: RD2022070103). M.B. and M.P.J. are financially supported by the European Union's Horizon 2020 research and innovation program under ERC advanced grant (grant agreement No 832237, SubNano). He also acknowledges the Centre de Calcul Intensif d'Aix-Marseille. O.I. acknowledges support from the National Science Foundation (NSF) CHE-2154447. O.I. acknowledges Extreme Science and Engineering Discovery Environment (XSEDE) Award CHE200122, which is supported by NSF Grant Number ACI-1053575. C.W. acknowledges funding support from the National Key R&D Program of China (2021YFA1502500), the National Natural Science Foundation of China (22071207, 22121001, 21721001, and 22003051), NFFTBS (no. J1310024), and the Fundamental Research Funds for the Central Universities (nos. 20720220128 and 20720220011).